\documentclass[final,5p,times,twocolumn]{elsarticle} 

\usepackage{lineno,hyperref}
\modulolinenumbers[5]
\journal{Journal of \LaTeX\ Templates}

\begin{document}

\begin{frontmatter}

\title{Large area LaBr3:Ce crystals read by SiPM arrays with improved
timing and temperature gain drift control}
%% \tnotetext[mytitlenote]{Fully documented templates are available in the elsarticle package on \href{http://www.ctan.org/tex-archive/macros/latex/contrib/elsarticle}{CTAN}.}


\author[ist1]{M.~Bonesini\corref{cor}}
\ead{maurizio.bonesini@unimib.it}
\author[ist1]{R.~Benocci}
\author[ist1]{R.~Bertoni}
\author[ist2,ist3]{A.~Menegolli}
\author[ist2]{M.~Prata}
\author[ist2]{M.~Rossella}
\author[ist2,ist3]{R.~Rossini}
\cortext[cor]{Corresponding author}

\address[ist1]{University and Sezione INFN Milano Bicocca, Piazza Scienza 3, Milano, Italy}
\address[ist2]{University of Pavia, via A. Bassi 6, Pavia, Italy}
\address[ist3]{Sezione INFN Pavia, via A. Bassi 6, Pavia, Italy}
%% \author{Elsevier\fnref{myfootnote}}
%% \address{Radarweg 29, Amsterdam}
%% \fntext[myfootnote]{Since 1880.}

%% \author[mymainaddress,mysecondaryaddress]{Elsevier Inc}
%% \ead[url]{www.elsevier.com}

%% \author[mysecondaryaddress]{Global Customer Service\corref{mycorrespondingauthor}}
%%\cortext[mycorrespondingauthor]{Corresponding author}
%%\ead{support@elsevier.com}

%% \address[mymainaddress]{1600 John F Kennedy Boulevard, Philadelphia}
%% \address[mysecondaryaddress]{360 Park Avenue South, New York}

\begin{abstract}
Compact X-rays detectors made of  1/2" or 1" :LaBr$_3$:Ce crystals of cubic 
shape with SiPM array readout have been developed for the FAMU experiment at
RIKEN-RAL. The aim is a precise measurement of the proton Zemach radius with 
incoming muons. Additional applications may be found in medical physics, such 
as PET, homeland security and gamma-ray astronomy. Due to the high photon yield of LaBr$_3$:Ce it was possible  to use a simple readout scheme based on 
CAEN V1730  digitizers.
Detectors, using Hamamatsu S13361 or S14161 arrays, have good FWHM energy 
	resolutions up to $3 \% $ ($8 \%$) at the $Cs^{137}$ ($Co^{57}$) 
	peak, comparing well with the best 
results obtained with a photomultiplier's  readout. 
Detailed studies were performed to correct 
online the drift with temperature of SiPM gain and to reduce the 
risetime/falltime of detectors' signals, that increased going from 1/2" to 1"
detectors, due to the larger capacity of the used SiPM arrays.

\end{abstract}


\end{frontmatter}

%%%\linenumbers

\section{Introduction}
LaBr3:Ce crystals have been introduced for radiation imaging in
medical physics, with a readout based on photomultipliers (PMTs) 
or on Silicon Photomultipliers (SiPM).
An R$\&$D was pursued with 1/2" and 1" LaBr3:Ce crystals, from different 
producers, to realize compact large area detectors (up to some cm$^2$ area) 
with a SiPM array readout. The aim was to obtain
high light yields, good energy resolution, good detector linearity
and fast time response for low-energy X-rays. A straightforward application was
found inside the FAMU ({\bf F}isica degli {\bf A}tomi {\bf Mu}onici) 
project \cite{famu} at the RIKEN-RAL muon facility. 
Its scope is a precise measure of the proton Zemach radius to contribute to
the assessment of the
so-called "proton radius puzzle", triggered by the recent measure of the proton charge 
radius with incoming muons at PSI \cite{antognini}.
In FAMU, the detection of characteristic X-rays around 130 keV, with a
short falltime of the output signal ($\leq 300-400 $ ns), is needed \footnote{ 
detectors' requirements and preliminary performances in beam, for 
detection of X-rays around 100 keV,  
were already shown in references \cite{famu} and  \cite{bonesini1}}. 
Other applications may be foreseen in medical physics, such as PET, 
gamma-ray astronomy and homeland security. For these  
good FWHM energy resolution
at higher energies ($\sim 500-600$ keV) is needed instead. 

The  drift of SiPM arrays gain  with temperature is a limiting factor,
giving a major deterioration of the FWHM  detector's energy resolution.
In addition, for large detectors (area $\sim$ 1'' square) the increment of the
SiPM array capacity increases the falltime, as compared
to what obtained with 1/2" detectors. 
To solve the first problem, a custom NIM module, based on CAEN A7585 digital 
power supply chips, was developed.
Test results of the correction of gain drift with temperature
for 1" SiPM arrays from Hamamatsu are reported. Previous results for
1/2"  arrays from Hamamatsu, Sensl and Avansid were already 
presented in references
\cite{bonesini1} and \cite{psd12}. 
At the $^{137}$ Cs ($^{57}$Co) peak,  an energy resolution  better
than  $ 3 \% (8 \%)$ was obtained for a typical 1" LaBr$_3$:Ce crystal, using
Hamamatsu S14161 arrays. This  compares well with best available results
obtained with a PMT. 
For the second problem (``timing'') different solutions were studied from
hybrid ganging of SiPM powering to the adoption of zero pole circuits, with or
without amplification. All solutions, while giving a sensible reduction of 
timing: up to a factor two, had the drawback of an  increase of the 
FWHM energy resolution: higher in the case of hybrid ganging. 
A good compromise was obtained by retaining the standard parallel ganging for
the powering of 
the cells of the SiPM arrays, with a suitable zero pole circuit, while 
improving the  used overvoltage by 1-2 V (at the limit of the operating 
conditions) to compensate for the signal decrease. 

\section{1" Crystal detectors with SiPM array readout}

The LaBr$_3$ crystals and the readout SiPM arrays  are mounted inside 
an ABS holder, realized with a 3D printer, as reported
in reference \cite{detec}. A parallel ganging for SiPM 
powering was used. The SiPM temperature is
monitored for an online correction of the operating voltage, via a TMP37 
temperature sensor.
Laboratory tests were done putting the detector under test
inside a Memmert IPV-30 climatic chamber, where
the temperature could be stabilized with a precision of $\sim$ 0.1 $^{\circ}$C.
Detectors were
powered at their
nominal operating voltage $V_{op}$.
Different exempt X-rays sources (Cd$^{109}$, Co $^{57}$, Ba$^{133}$, Na$^{22}$,
Cs$^{137}$, Mn$^{54}$) were used for calibration.
The summed analogue signal from the cells of a SiPM array is directly fed into a CAEN V1730
fast digitizer (500 MHz bandwidth, 14 bit resolution) and is acquired by a custom
developed DAQ system \cite{soldani}. Produced n-tuples are analyzed by the
ROOT package.
\begin{figure}[htbp] % figures (and tables) should go top or bottom of^M
                    % the page where they are first cited or in^M
                    % subsequent pages^M
\centering
\includegraphics[width=0.44\textwidth]{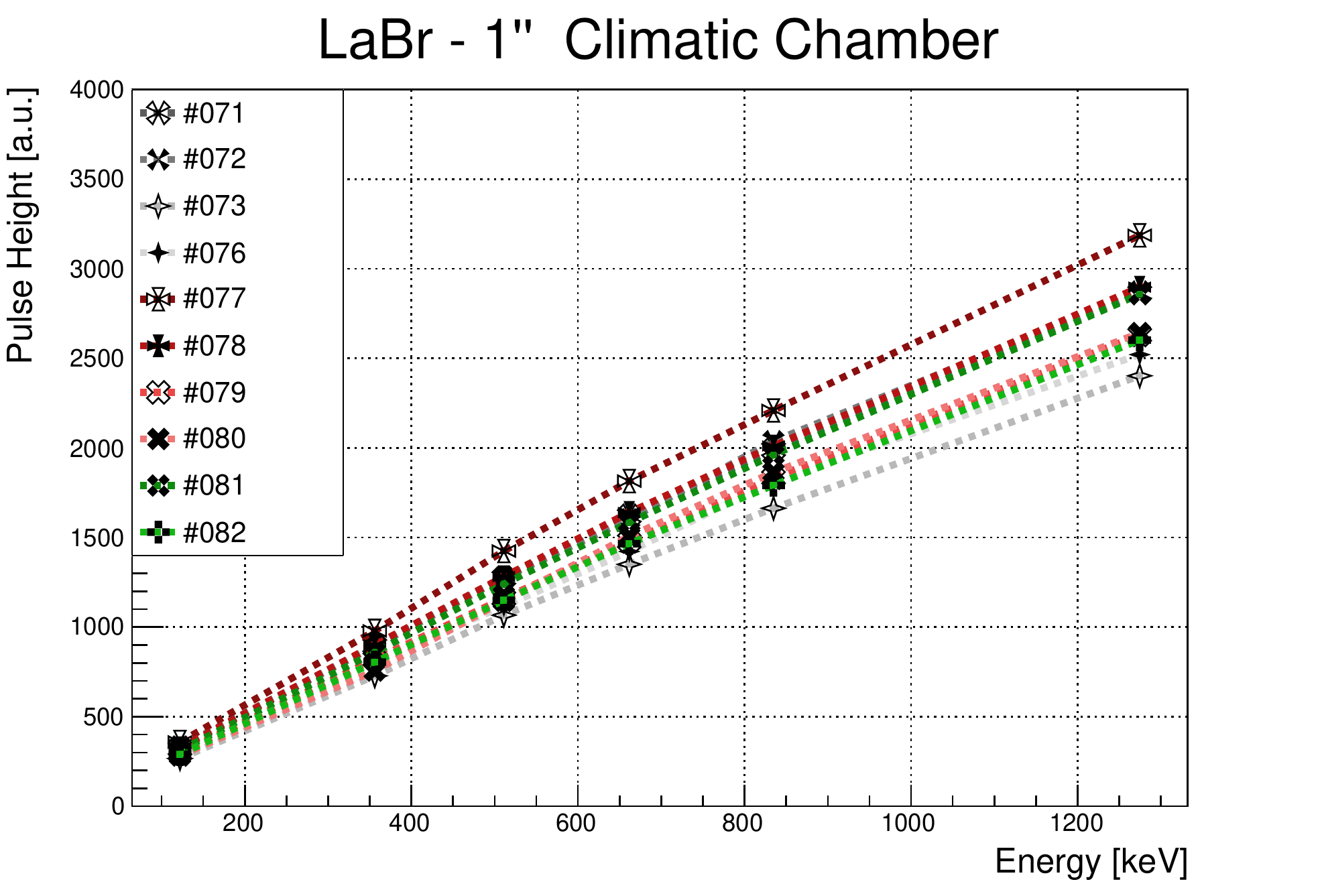}
\includegraphics[width=0.44\textwidth]{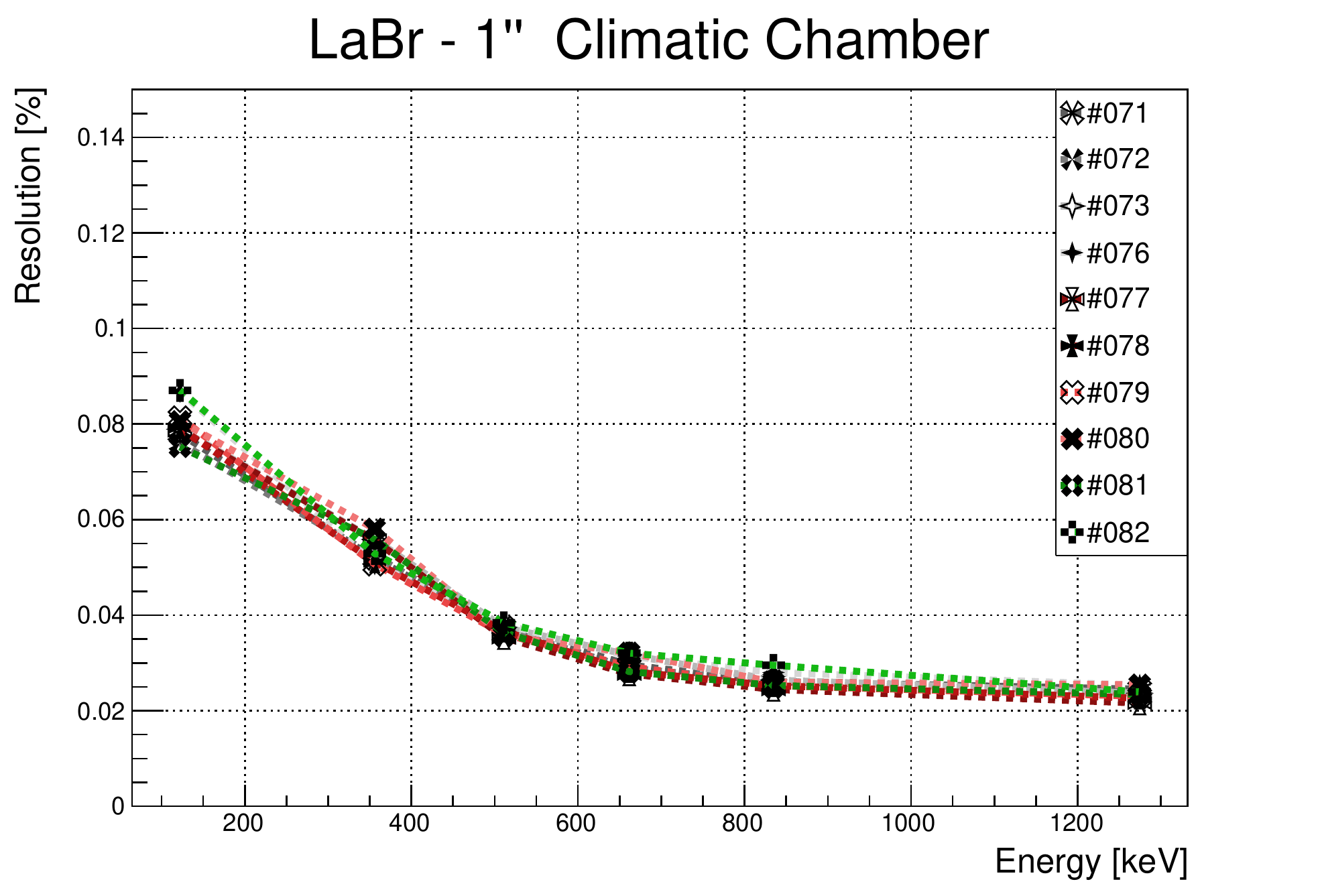}
\caption{Top panel: linearity for a sample of  1" detectors. Bottom panel: 
FWHM energy resolution for the same 1" detectors. Temperature control is applied.
}
\label{fig-res}
\end{figure}
Figure \ref{fig-res} shows  results for linearity and FWHM energy resolution
obtained for  a sample of 1" crystals 
at the reference temperature  of $25 ^{\circ}C$ inside the climatic chamber.
Results for 1/2" crystals has been previously shown in references \cite{bonesini1} and 
\cite{detec}.

\section{Correction of gain drift with temperature}

\begin{figure}[htbp] % figures (and tables) should go top or bottom of^M
                    % the page where they are first cited or in^M
                    % subsequent pages^M
\centering
\vskip -1cm
\includegraphics[width=0.23\textwidth]{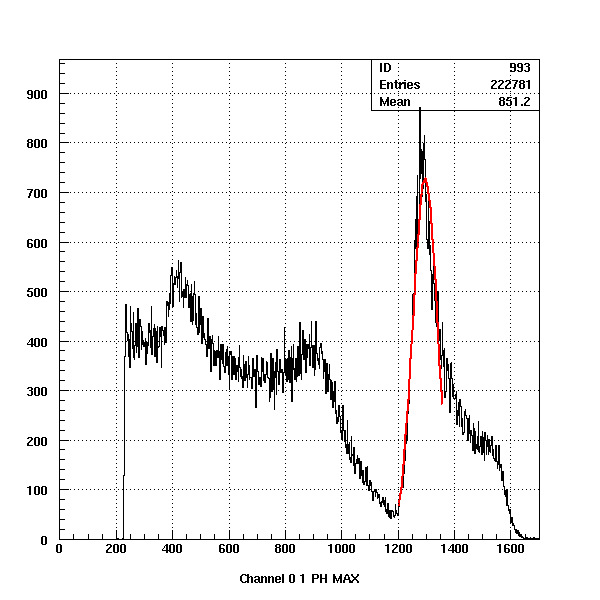}
\includegraphics[width=0.24\textwidth]{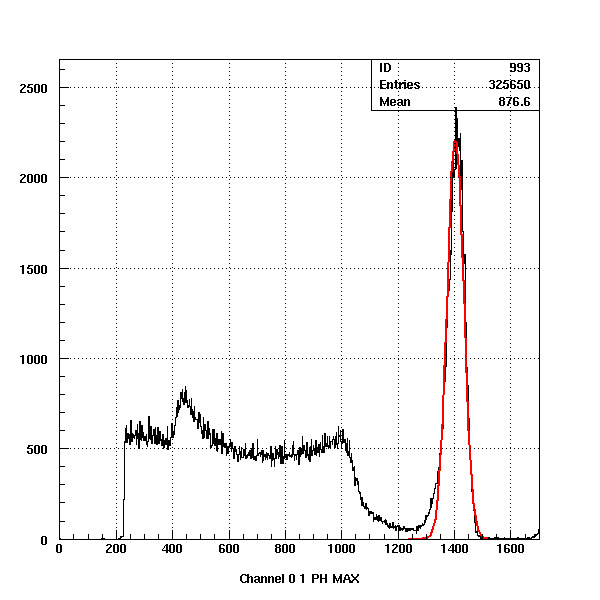}
	\caption{ Cs$^{137}$ spectra recorded by a LaBr3:Ce 1" 
	detector read by an Hamamatsu 14461 SiPM  array during 
	a temperature scan between 20 $^{\circ}$C and
	30 $^{\circ}$C, inside a climatic chamber. Left panel: is without 
	temperature correction, 
	Right panel: is with  online temperature correction. Fits to the 662
	keV photo peak, with a simple gaussian , are shown}.
\label{fig:crys1}
\end{figure}

The SiPM's gain drifts significantly as a function of temperature.
This feature prevents their use in conditions with a changing
temperature, as homeland security and military
applications.
The response of a typical 1" detector to a $^{137}$Cs source 
in the IPV-30  climatic chamber, where the
temperature is varying from 20 to 30 $^{\circ}$C, is shown in the left panel of 
figure \ref{fig:crys1}. Without the online correction, the resolution of 
the photo peak at 662 keV is 
sensibly degraded. The response, with the online correction for temperature 
(see below for further details) is shown instead in the right panel of the
same figure, where no degradation of the $^{137}$Cs photo-peak is seen. 
To overcome this problem, SiPMs must be operated at a fixed gain if
temperature changes. Thus the operating voltage $V_{op}=V_{bd} + \Delta V$,
with $V_{bd}$ breakdown voltage and $\Delta V$ overvoltage, must be modified
as a function of temperature according to:
$$ V_{bd}(T) = V_{bd}(T_{ref}) \times (1 + \beta (T-T_{ref})) $$
with T working temperature, $T_{ref}$ reference temperature (typically 25$^{\circ}$ C)
and $\beta =\Delta V_{bd}/ \Delta T$ the differential value of the breakdown
voltage to temperature. $\beta$, as explained in references \cite{dinu} and \cite{otte}, 
is independent of temperature. 

For an online hardware correction,
a custom NIM module, based on CAEN A7585D digital
power supplies, with temperature feedback  was developed. The SiPM temperature is monitored via
a TMP37 sensor from Analog Devices \footnote{$\pm 2^{\circ}$C accuracy over
temperature, $\pm 0.5^{\circ}$C linearity}, mounted on the PCB where the array
socket is soldered. A 3.5 mm stereo jack cable connects the sensor to the 
custom NIM module for online temperature correction. 

\begin{figure}[htbp] % figures (and tables) should go top or bottom of^M
                    % the page where they are first cited or in^M
                    % subsequent pages^M
\centering
\includegraphics[width=0.50\textwidth]{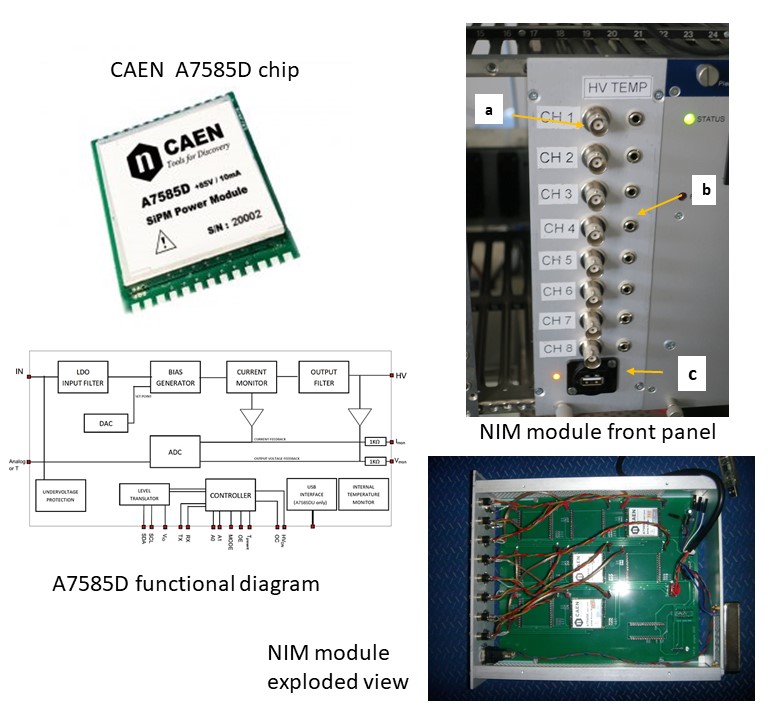}
\vskip -0.5cm
\caption{Left panel - top: image of one CAEN A7585D module; bottom -
	A7585D functional layout (courtesy of CAEN srl).
        Right panel -top: front panel of one custom NIM modules: a) is the BNC
         connector for a channel HV, b) the 3.5 mm jack stereo for the
         connection to a temperature TMP37 sensor, c) the USB interface
         connection; bottom: exploded view of the custom NIM module.}
\label{fig-module}
\end{figure}

Up to eight channels may be powered
by a single 2-slots NIM module, as shown in figure \ref{fig-module}.
The control of the NIM  module was implemented  via either
a FDTI USB-I2C converter or an  Arduino Nano chip.
Three modules (with which up to 24 channels may be powered) were realized 
and are linkable in daisy chain, via the I2C protocol.

\begin{figure}[htbp] % figures (and tables) should go top or bottom of^M
                    % the page where they are first cited or in^M
                    % subsequent pages^M
\centering
\includegraphics[width=0.49\textwidth]{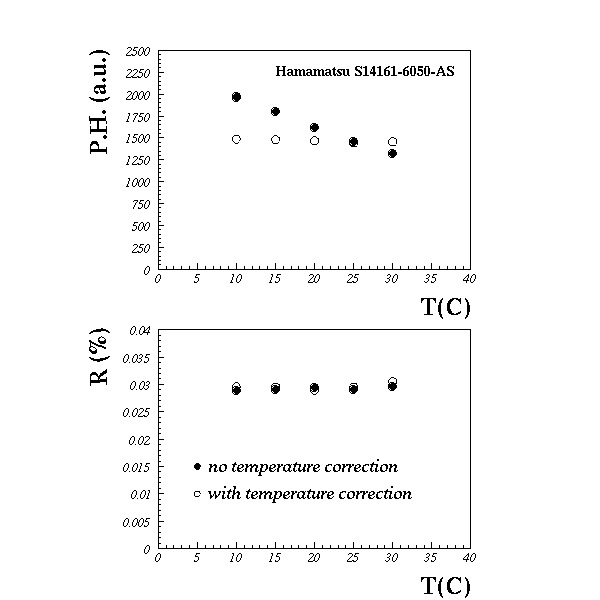}
	\caption{Dependence of the photo peak position (top) and the FWHM 
	energy resolution at 662 keV for a typical 1" 
detector, with and without temperature correction.
}
\label{fig-crys}
\end{figure}

One inch LaBr$_{3}$:Ce crystals, read by Hamamatsu S14161 SiPM arrays, 
were then put inside the IPV-30 climatic chamber 
for measurement. 
With a $^{137}$Cs exempt source, tests were done with and without 
the online
temperature corrections. 
Results on the dependence of the photo peak at 662 keV and the FWHM energy
resolution, 
as a function of temperature, are 
reported in figure \ref{fig-crys}. The same crystal is used in all the tests.
After correction, the variation in the $^{137}$Cs photo-peak position 
(up to 33 $\%$ in the range 10-30 $^{\circ}$C) was reduced to $\sim  7 \% $.
Similar results with 1/2" LaBr$_{3}$:Ce crystals, read by Hamamatsu, Advansid
or SENSL  SiPM arrays, were previously reported in reference \cite{detec} 
in the range  10-45 $^{\circ}$C.
%% While Hamamatsu and Advansid SiPM arrays showed a larger response variation 
%% (up to 70 $\%$), the effect was smaller for SENSL devices ($\sim 30 \%$).
After the temperature 
correction, in all cases the variation was reduced to $\sim 5 \%$. 
The online temperature correction for gains seems to work better for
the smaller 1/2" SiPM arrays.
As shown in the bottom panel of figure \ref{fig-crys}, 
the effect on the FWHM energy resolution 
is within 1 per mille and is compatible with a zero dependence. 
Our results confirms
previous results as published in references \cite{Moszynski} and 
\cite{Hou},  both obtained with a PMT readout.
\vskip -0.4cm

\section{Improving timing properties}

\begin{figure}[htbp] % figures (and tables) should go top or bottom of^M
                    % the page where they are first cited or in^M
                    % subsequent pages^M
\vskip -2cm
\centering
\includegraphics[width=0.59\textwidth]{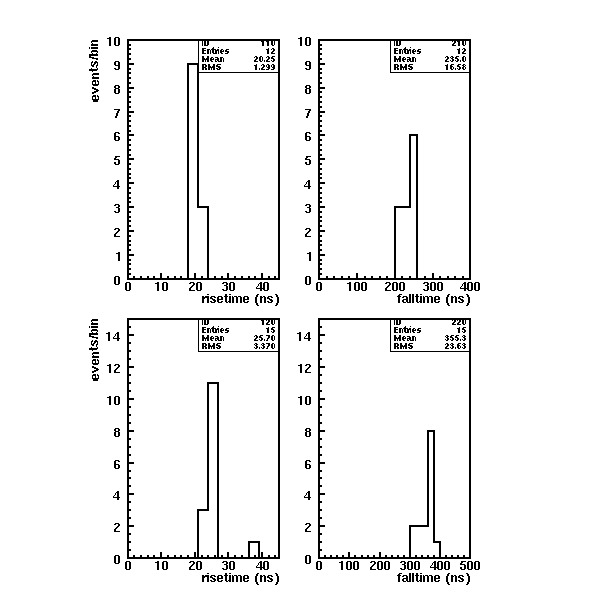}
	\caption{ Distributions of the risetime (20-80 $\%$) and falltime
	for 1/2" LaBr3:Ce crystals read by Hamamatsu S13361-3050-AS
	(top panel) or S14161-3050-AS (bottom panel) SiPM arrays. The 
	difference in risetime/falltime
        between upper and lower plots, is due to the capacity of the used 
	arrays:
        320 pF vs 500 pF. A standard parallel ganging is used.
}
\label{fig-time2}
\end{figure}
\begin{figure}[htbp] % figures (and tables) should go top or bottom of^M
                    % the page where they are first cited or in^M
                    % subsequent pages^M
\vskip -1cm
\centering
\includegraphics[width=0.49\textwidth]{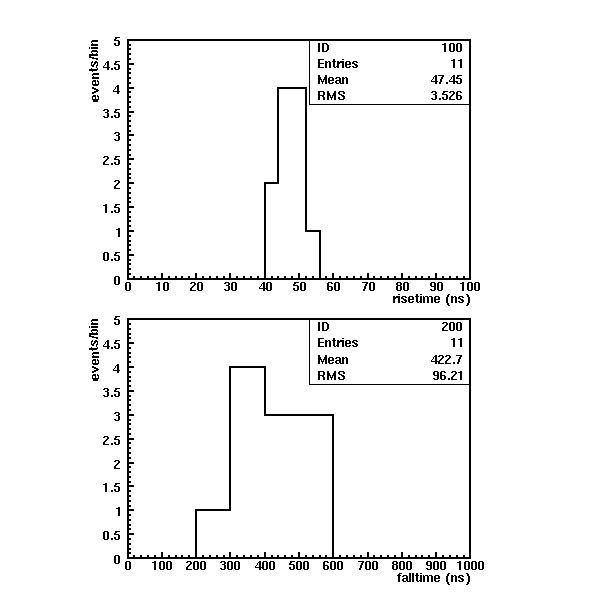}
	\caption{ Distributions of the risetime (20-80 $\%$) and falltime
	for 1" LaBr3:Ce crystals read by S14161-6050-AS Hamamatsu  
	SiPM arrays. A standard parallel ganging is used.
}
\label{fig-time}
\end{figure}
Different schemes may be used for powering the cells of a SiPM array,
from parallel ganging, to series ganging or to hybrid ganging, as shown
in reference \cite{colazzuol}. 
The standard parallel ganging of SiPMs inside an array provides a good
FWHM energy resolution: up to $\sim 3 \%$, comparable to what 
obtained with a conventional PMT readout \cite{famu1}, but with poor timing
properties  for larger SiPM arrays (e.g. 1 inch array).

\begin{figure}[htbp] % figures (and tables) should go top or bottom of^M
                    % the page where they are first cited or in^M
                    % subsequent pages^M
\centering
\includegraphics[width=0.5\textwidth]{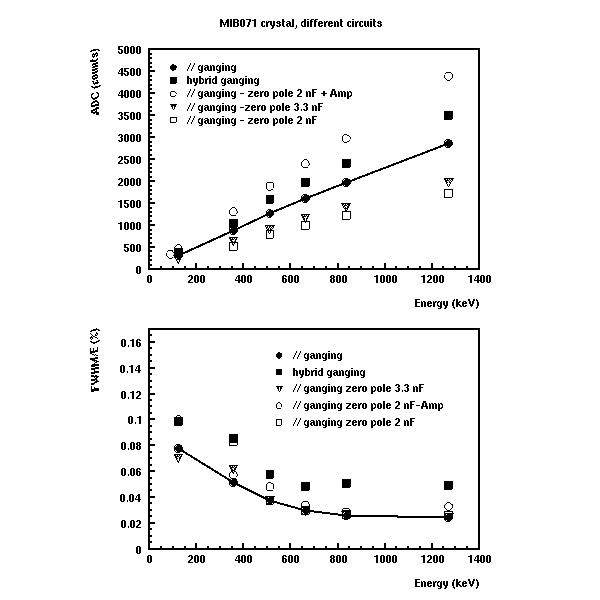}
\caption{Effects on linearity and FWHM energy resolution for a typical 1" 
LaBr3:Ce crystals with different readout circuits. The line connects points
for the standard parallel ganging of the SiPM cells of the array. 
}
\label{fig-crys1}
\end{figure}

Increasing the dimensions of the SiPM arrays, to match the growing
detector's window area, timing properties worsen, due to 
the increased array capacity. As an example, for S14161 Hamamatsu SiPM arrays
the capacity increases from 500 pF to 2 nF, going from 1/2" to 1" size.
Figures \ref{fig-time2} 
and \ref{fig-time} report
the distribution of risetime and falltime for a sample of 1/2"  and 1" 
detectors. A clear increase in both risetime and falltime is seen.  
Different solutions were studied for 1"  detectors, from hybrid ganging to
the use of a zero pole circuit with or without an amplifier stage, to reduce
timing. 
A reduction of risetime/falltime up to a factor 2 was obtained, at the
cost of an increased FWHM energy resolution. Results for a typical detector
are reported in table \ref{tab1} and figure \ref{fig-crys1}.

%%\begin{figure}[htbp] % figures (and tables) should go top or bottom of^M
%%                    % the page where they are first cited or in^M
%%                    % subsequent pages^M
%%\centering
%%\includegraphics[width=0.3\textwidth]{fig_hybrid_1.png}
%%\includegraphics[width=0.3\textwidth, angle=-90]{fig_hybrid_2.pdf}
%%\caption{Top: image of the PCB with hybrid ganging for SiPM cells of a S14161 
%%	array.Bottom: layout of the same PCB.}
%%\label{fig-hybrid}
%%\end{figure}

\begin{table*}[htb]
\smallskip
\vskip -2cm
\caption{Timing and energy resolution for a typical (MIB071) 1" LaBr3:Ce
	crystal, with different solutions for the electronic readout}
\centering
\begin{tabular}{|c|c|c|c|c|c|}
\hline
	circuit  &  V$_{op} (V) $ & risetime   & falltime  & resolution($\%)$ & 
	 resolution ($ \%)$ \\ 
	&          & (20-80$\%$) (ns)  & (10-90$\%$) (ns) & $@ Co^{57} $ & 
	$@ Cs^{137} $  \\ \hline
	// ganging & 41.82 & $ 46.6\pm 5.3 $ &  $ 293.3 \pm 43.4$ &7.78 &  2.96 
	  \\ \hline
	hybrid ganging & 41.82 & $16.1 \pm 2.4$ & $176.8 \pm 29.0 $ & 9.58 &
	6.08  \\ \hline
	0-pole 3.3 nF & 43.02 & $ 50.0 \pm 16.5 $ & $145.7 \pm 33.2 $ & 7.0 &  
	2.94  \\
	\hline
	0-pole 2 nF   &  43.02 & $36.0 \pm 7.9 $ & $ 123.4 \pm 21.7 $ &  - & 
	2.99  \\ \hline
	0-pole 2 nF + Amp & 40.82 & $29.4 \pm 12.6 $ & $170.8 \pm 48.7 $ & 
	10.0 & 3.40  \\    \hline
\label{tab1}
\end{tabular}
\end{table*}
A good compromise is obtained by using conventional parallel ganging, with 
a 2 or 3.3 nF zero pole circuit increasing the overvoltage $\Delta$ V to compensate
for the reduction of the signal amplitude \footnote{
Rising the overvoltage by 1-2 V, a 15-30 $ \%$ increase in
the dark count rate for Hamamatsu S14160 SiPM may be expected, as shown in reference
\cite{Li}.  As dark counts are undistinguishable from photons events, a 
degradation in the energy
resolution may thus result.  However, we had no evidence of such an effect,
with a pole zero circuit + increased overvoltage of SiPM}. 

\section{Conclusions}
Good FWHM energy resolutions are obtained with  1" LaBr$_{3}$  crystals
read by Hamamatsu S14161-6050-AS SiPM arrays, both at the $^{137}$Cs peak
and at lower energy at the $^{57}$Co peak. Resolutions up to 3 $\%$ and
$ 8 \%$ were obtained in the two cases. With 1" detectors there is an increase 
of signal risetime and falltime, due to the increased 
capacity of the used SiPM arrays, as respect to 1/2" ones. 
To reduce the falltime of the 1" detectors,
different solutions were studied. A simple solution based on a zero pole
circuit and a greater overvoltage for  the used SiPM arrays 
gives a reasonable reduction of
falltime/risetime at the expense of a minimal increase of FWHM energy 
resolution. Solutions based on the use of hybrid ganging show a sensible 
increase of the detectors' energy resolution and were discarded.

\end{document}